# Navigating the Topology of 2x2 Games:
# An Introductory Note on Payoff Families, Normalization, and Natural Order

Bryan Bruns
bryanbruns@bryanbruns.com

*The Robinson-Goforth topology of swaps in adjoining payoffs elegantly arranges 2x2 ordinal games in accordance with important properties including symmetry, number of dominant strategies and Nash Equilibria, and alignment of interests. Adding payoff families based on Nash Equilibria illustrates an additional aspect of this order and aids visualization of the topology. Making ties through half-swaps not only creates simpler games within the topology, but, in reverse, breaking ties shows the evolution of preferences, yielding a natural ordering for the topology of 2x2 games with ties. An ordinal game not only represents an equivalence class of games with real values, but also a discrete equivalent of the normalized version of those games. The topology provides coordinates which could be used to identify related games in a semantic web ontology and facilitate comparative analysis of agent-based simulations and other research in game theory, as well as charting relationships and potential moves between games as a tool for institutional analysis and design.*

The topology of 2x2 games (Robinson and Goforth 2005) offers an elegant tool for understanding relationships among ordinal games and their transformations. Game theory research has concentrated on Prisoner's Dilemma and a few other symmetric games, although most of the possible games are asymmetric, and most possible ordinal games have ties. The topology helps show both the uniqueness of Prisoner's Dilemma, Chicken, and the games closely related to them, and how they fit into the much larger and more diverse landscape of games, many of which are assurance (Stag Hunt/coordination) games or have a harmonious cooperative (win-win) outcome, and even more of which have asymmetric outcomes. This paper briefly introduces the topology of 2x2 games and outlines how payoff families, normalization of payoff values, and natural ordering of preference structures aid understanding relationships among games in the topology.

The topology organizes games by closeness in terms of swaps in adjoining payoffs, revealing patterns in symmetry, number of dominant strategies and Nash Equilibria, and alignment of interests. Adding payoff families based on Nash Equilibria reveals another aspect of order in the topology (Bruns 2010a). Robinson and Goforth plot payoffs on order graphs to efficiently display symmetries and mixes of interests (Schelling 1960; Greenberg 1990; Robinson and Goforth 2005) as shown in their "periodic table" of 2x2 ordinal games.[1] A numeric display in normal form with additional visualization provides a more accessible version of the topology, which may make the topology easier to learn and use, including for analysis of potential moves between games (Bruns 2010c). Making ties between adjoining payoffs extends the topology to non-strict games (Robinson, Goforth, and Cargill 2007) and can be arranged according to the distinct classes of preference structures for ties identified by Fraser and Kilgour (1986). The reverse process of breaking ties to create different payoffs indicates a natural ordering for the topology, and establishes a grid of coordinates useful for identifying related games, including normalized versions of games with real payoffs (Bruns 2010b).

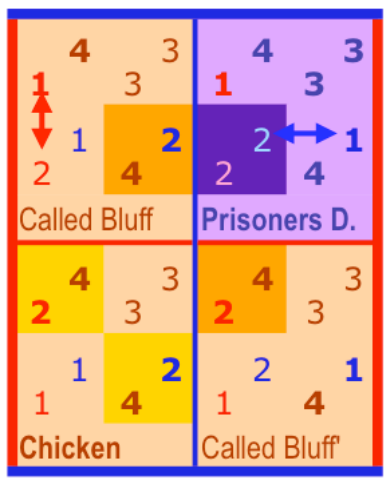

Figure 1. Tile of 1↔2 swaps

*Swaps.* In the Robinson-Goforth topology of the payoff space of 2x2 games, swaps in the two lowest payoffs (1↔2) form *tiles* of four games, defining the games that are closest to each other. Swaps in middle payoffs (2↔3) and further low and mid swaps create a *layer* of nine tiles, and thirty-six games, which is a torus. Swaps in high payoffs (3↔4) create games on other layers, leading to the full topology of strict (no ties) ordinal games (Plate 1).

*Layers and payoff families.* Symmetric games lie on a diagonal from lower left (southwest) to upper right (northeast) (Plate 2a). In each layer, the lowest three rows have dominant strategies for the row player, and the left three columns have dominant strategies for the column player (Plate 2b). The regions with dominant strategies have a single Nash Equilibrium, while those without dominant strategies have either no Nash Equilibria, for the Cyclic games; or two, for the Stag Hunt and Battle of the Sexes families of games.

Payoff families based on Nash Equilibria provide a way of categorizing games by outcomes (Plate 1 and Plate 2c) and further revealing structure in the topology. Alignment of the two highest payoffs distinguishes layers (as shown by the payoff patterns inset in Plate 2a). Layer One, with the highest payoffs diagonally opposite, contains the symmetric games most studied by researchers: Prisoner's Dilemma, Chicken (often analyzed in evolutionary biology as Hawk and Dove strategies), and Battle of the Sexes, a variant of which is also called Hero (Rapoport 1967). Asymmetric variants of Battle of the Sexes still have identical Nash Equilibria, while their *improper* cousins mix 4,3 and 4,2 equilibria, combining Battle of the Sexes and Chicken outcomes. Layer One also contains the Prisoner's Delight game (Binmore 2007) and its variants, with second-best outcomes for each player at the Nash Equilibrium.

[1] Available at www.cs.laurentian.ca/dgoforth/home.html

Layer Three has the two highest (*win-win)* payoffs in the same cell, forming mostly *harmonious* games with a single equilibrium based on dominant strategies for one or both players. The four *proper* Stag Hunt games, have 4,4 and 3,3 equilibria, bordered by games that mix 4,4 with 3,2 equilibria and, in one case a 2,2 equilibrium. Layer 3 games have been labeled as "no conflict" or "boring." However, as Robinson and Goforth point out, even many of the games with a single win-win equilibrium are games of mixed interests (Plate 2f). "No conflict" is a misnomer. The Stag Hunt (assurance/coordination) games with two different Nash Equilibria are particularly interesting for social theory (Sen 1967; Kollock 1998; Skyrms 2004).

Layers Two and Four are mirror images of each other that differ only by switching the positions of row and column players, part of the symmetric structure of the topology along the axis created by the symmetric games. Layers Two and Four include the well-known category of Cyclic games. Somewhat surprisingly, the largest subfamily, on these layers and overall, is composed of Samaritan games where a player following a dominant strategy gets their second-ranked outcome (exemplified by Buchanan's (1977; Schmidtchen 2002) model of Active Samaritan's Dilemma, Game 262).

Next to the Samaritan games are Unfair games, where a dominant strategy would lead a player to their third-ranked outcome. It should be noted these outcomes would result from playing dominant strategies based on a narrow, short-sighted rationality. Nash Equilibria need not match results from actual play, given bounded rationality, heuristics, learning, and other strategic factors, especially if communication is possible and play is repeated, (or if play begins from a particular starting point, as with the game of Samson and Delilah (Game 213) in Bram's (1994) *Theory of Moves*).

Layers Two and Four include asymmetric siblings and cousins to Prisoner's Dilemma with poor Pareto-inefficient Nash Equilibria, forming the Prisoner's Dilemma Family identified by Robinson and Goforth. This family can extend to include the adjoining Tragic games that also have poor 3,2 outcomes, but lack even the possibility of a Pareto-superior outcome. This layer also has additional games with 3,3 equilibria, part of a family of Second Best games.

The 12 symmetric games plus the 66 games above or below that axis of symmetry (for switching positions of row and column players) compose the 78 unique strict ordinal games identified by Rapoport and colleagues (Rapoport and Guyer 1966; Rapoport, Guyer, and Gordon 1976). The topology includes key distinctions from that earlier taxonomy (Plate 2h), including the presence of win-win (4,4 or "no conflict") outcomes, two Nash Equilibria or none, and a Pareto-deficient equilibrium (though only identified by them for Prisoner's Dilemma and not for its siblings and cousins). The topology allows overlapping categories, rather than the Linnaean hierarchy attempted by Rapoport and colleagues, and does not depend on somewhat debatable judgments about threats and responses. Another typology and ordering of 2x2 games was developed by Brams (1994). Plotting this on the topology (Plate 2h) shows that his categories, and concepts concerning non-myopic equilibria, mainly cover games on Layers Two and Four (and he ignored Layer Three).

The topology locates the symmetric games within the much larger number of asymmetric games (Plate 2a), puts Prisoner's Dilemma and other games with Pareto-inferior and poor outcomes into the context of the games with better, but often biased, outcomes at Nash Equilibria (Plate 2c), and shows how games of pure conflict and of pure cooperation are greatly outnumbered by games with mixed interests (Plate 2f).

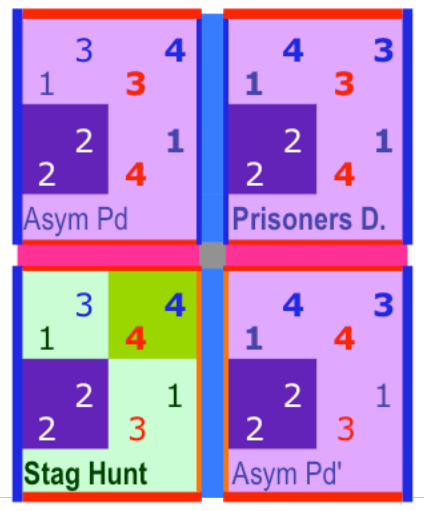

Figure 2. 3↔4 swaps

*Hotspots and pipes.* Swaps in the two highest payoffs link layers, with all the games in each row or column *slice* of tiles moving to the same layer, for row or column swaps respectively (2d). In *hotspots*, high swaps for both row and column link two tiles on two layers. *Pipes* weave together four tiles on four layers. Links in pipes and hotspots can be followed more easily by separating movement into three components, for layer, tile, and game, corresponding to changes in the location of the four, three and two payoff values. Swapping a four changes layer, swapping a three changes tile, and swapping a one or two moves within a tile. (These components can also map swaps between non-adjoining payoffs, as in Buchanan's two versions of Samaritan's Dilemma.) For institutional analysis and design, swaps map how changes in payoff ranking transform incentive structures, for example the two swaps, one for each player, that transform Prisoner's Dilemma into a Stag Hunt (Figure 2) or the single swaps that change Samaritan's Dilemma into a win-win game (Bruns 2010c).

*Ties.* Half-swaps make (or break) a tie between two adjoining payoffs, thus creating games in between those in a tile (Robinson, Goforth, and Cargill 2007) (see Plates 2g and 3), One example is the game midway between Prisoner's Dilemma and Chicken (Figure 3 and Plate 3a) (Oskamp 1971; Kilgour and Fraser 1988; Rapoport, Guyer, and Gordon 1976, 31). Combined swaps for both row and column create a game at the center of each tile, usually a simplified or idealized version of the games in the tile. Taken together, the games with ties for the two lowest payoffs form a simplified matrix of Tile Games (Plate 3c). Rousseau's story where all must cooperate to hunt a stag, but a player can get a hare regardless of what others do, corresponds (in the two-player version) to a game with ties for the middle payoffs, located diagonally between two tiles (Plate 3b). All ordinal games with ties can be seen as located between the games in the strict topology.

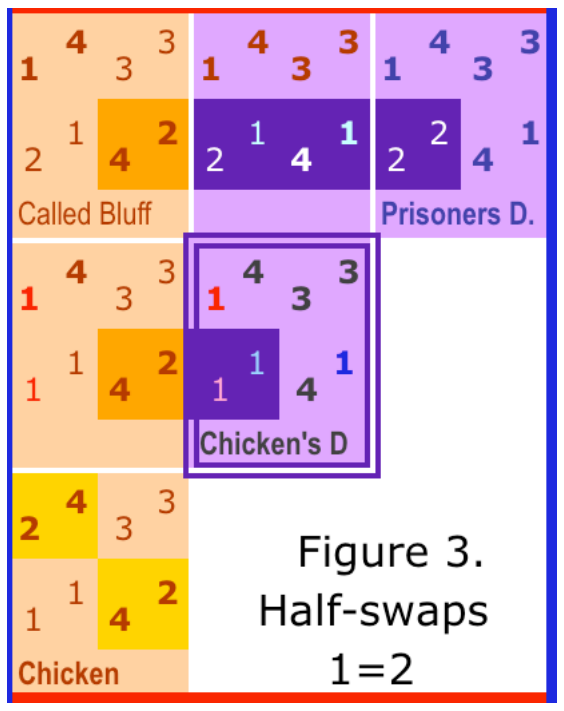

Figure 3. Half-swaps 1=2

Additional ties form simpler games, down to Utter Harmony, the simplest game where both have at least one preference (Plate 3d). Robinson et al. noted that the games with two pairs of tied preferences form archetypal games, including Zero Sum (or, more precisely, a fixed sum game). Despite some claims that people are unlikely to play mixed strategies in real life, simple games on the same matrix as Utter Harmony pose situations where mixed or synchronized strategies would be very advantageous in repeated play.

*Natural order.* Robinson, Goforth and Cargill (2007) arranged their table of the number of unique non-strict ordinal games by the preference classes identified by Fraser and Kilgour (1986) resequenced to progress from the games with ties on the lowest payoffs to those with ties on higher payoffs (Robinson, Goforth, and Cargill 2007)(Plate 4b). Players may have two, three, or four distinct preferences (Plate 4a). For two preferences, there are three patterns: ties for the low rank, a pair of ties, or top rank. For three preferences, there may be ties for the low, middle, or high values. Contrary to Fraser and Kilgour's assertion (1986, 103) that "there is no natural order" for the non-strict games, the topology with ties, as resequenced by Robinson et al. shows a natural progression in numbers of preferences, leading from Utter Harmony (or the even simpler null game where neither player has a preference) at the southwest corner to Prisoner's Dilemma in the northeast corner (Plate 4a) (Bruns 2010b).

This natural ordering also implies that if one wants to follow the conventional arrangement (used by Robinson and Goforth within games) of putting higher values along the north and east axes, then their table should be flipped, to move Prisoner's Dilemma from the southwest to the northeast. However, for convenient display of game properties, including swaps in high payoffs for Prisoner's Dilemma and neighboring games, it is still convenient to scroll the (torus) display to put Prisoner's Dilemma near the center, as in Plate 1.

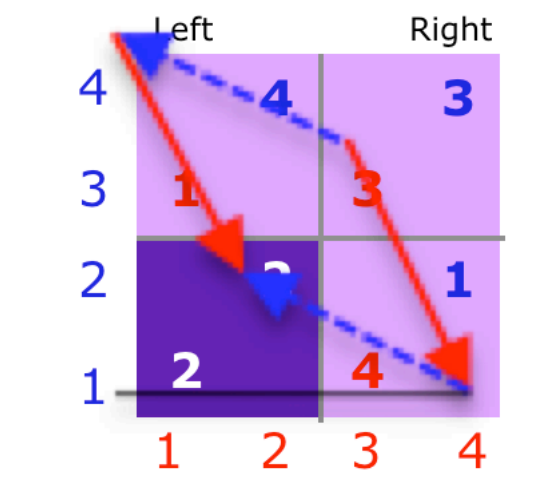

Figure 4. Order Graph

*Normalizing real games.* As discussed by Robinson and Goforth, each ordinal game represents an equivalence class of games with real number payoffs, although their simulation studies pointed out that games with the same ordinal structure behave differently and chaotically, if minor errors are allowed. The games with four discrete values from one to four may be seen as discrete equivalents to games with payoffs measured on interval or cardinal scales. Normalized versions of those games could be displayed on order graphs. The topology of ordinal games thus provides a set of coordinates for mapping the locations of games, and navigating between games through swaps or half-swaps in discrete ordinal payoffs, and through smaller changes in real values.

*Coordinates.* Robinson and Goforth's numeric indexing could be used for the matrices of games with ties (Plate 4a). In conjunction with Fraser and Kilgour's preference classes for ties (A-H), this would identify the location of specific games. Alternatively, "geographic" references (north-south-east-west) could be used, along with an index of preference classes. For a display arranged in the "natural order" with Prisoner's Dilemma in the far northeast corner (Plate 5), then Rousseau's Stag Hunt, with ties on the two middle preferences and a win-win outcome, would be game GG311, at the intersection of the $2_2$ preference class, in the southwest layer, on the northeast corner: (Game $2_2 2_2$swNW, as in Plate 4a).

Robinson and Goforth's convention for displaying games puts the column with Row's four to the right (east) and the row with Column's four up (north), so win-win (4,4) payoffs are in the upper-right (northeast) cell. Games with row's four to the west or column's four to the south could be shown as quadrants of a larger display, thereby including games equivalent by switching rows, columns, or both. Coordinates with the quadrant in this larger display would further specify games (implicitly, NE in Robinson and Goforth's convention). Research on heuristics and biases (Tversky and Kahneman 1990) found that responses to gains or losses are not symmetric, and similar differences may apply to otherwise equivalent games. It is worth noting that to the extent game payoffs occur randomly, and are not constrained to a small range of integer values, they will approximate the proportions in the topology of 144 strict ordinal 2x2 games (Simpson 2010).

Universal resource identifiers (URIs) for games could help identify research on equivalent or similar games and facilitate comparative analysis, for example findings from agent-based modeling, experiments, and other sources. For the semantic web (Berners-Lee et al. 2001), the best identifiers for games would be payoffs themselves, suitably encoded (a format such as game (a,d;b,e/c,f;d,g). The topology gives coordinates for locating games and metrics for how closely they are related. The topology thus provides the basis for an ontology of games, (in the computer science or information science sense of the term); more generally, the topology graphs a connected ontology in payoff space (ecologically) and joint preferences (cognitively).

*Extensions.* As with game theory generally, the topology can extend to multiple players or moves, and preferences measured on ratio or real scales (Plate 4d). In the payoff space, the topology can extend to payoffs that are fuzzy, probabilistic, uncertain, or dynamic.

*Conclusions.* The 2x2 games are only a small elementary enclave within a vast and complex multidimensional space of joint preferences and actions. Nevertheless, such games can aid understanding of institutional diversity (Ostrom 2005, 6), not only of strategies but also of incentive structures and how they might be transformed. Visualization of payoff families, ordering games with ties by preference classes, and normalization of real payoffs further illustrate how the topology arranges the connections between 2x2 games. Overall, the topology links 2x2 games in an elegant order, showing relationships among games and their properties, guiding navigation between games, and aiding analysis of potential transmutations between games as part of institutional analysis and design.

# 1. Topology of 2x2 Games with Payoff Families

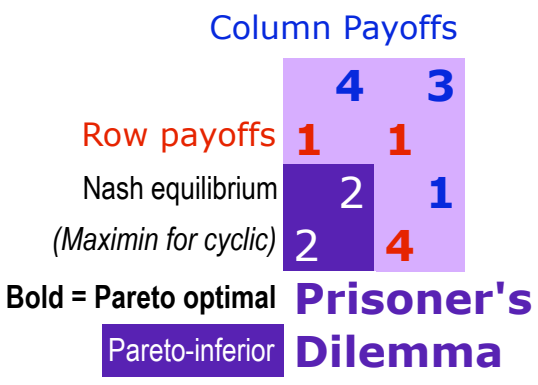

Adjacent games are neighbors by payoff swaps
1↔2 swaps form tiles of 4 games
2↔3 swaps link tiles into four layers
3↔4 swaps switch layers
Each layer is a torus, table is a torus
Layers scrolled to center Pd
Numbering: Layer-Row-Column Pd=111

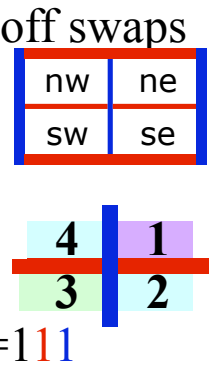

| | |
|---|---|
| 1. WIN-WIN 4-4 | Harmonious |
| | Stag Hunt |
| 2. BIASED 4-3 | Samaritan |
| Battles of the sexes | Self-serving |
| 3. SECOND BEST 3-3 | |
| 4. UNFAIR 4-2 | |
| 5. PD FAMILY | Pd 2-2 |
| Alibi 3-2 | Tragic 3-2 |
| 6. CYCLIC - (no Nash Equilibrium) | |

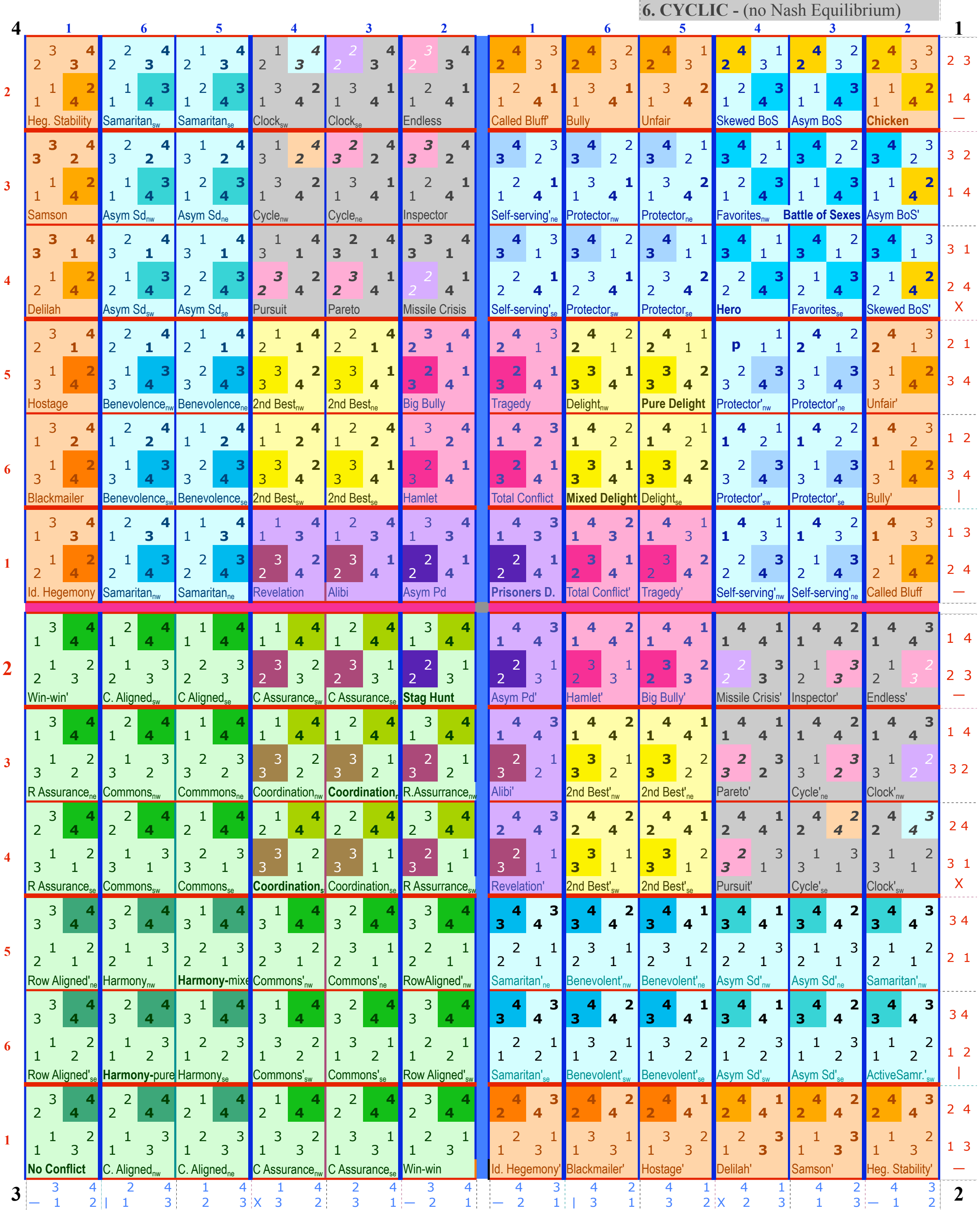

**To find a game:** Make ordinal 1<2<3<4
Put column with Row's 4 right; put row with
Column's 4 up. Then find layer,
Row's 4 up=Layers 2&3; down=1&4
Column's 4 left=Layers 1&2. Right=3&4
Find Row & Column payoffs
Game is at intersection.

Modified from Robinson and Goforth 2005
*The Topology of 2x2 Games: A New Periodic Table*
also see: www.cs.laurentian.ca/dgoforth/home.html

Available at www.bryanbruns.com/2x2chart.pdf

# 2. The Topology Arranges 2x2 Games in an Elegant Order

## a. Twelve Symmetric Games on Diagonal

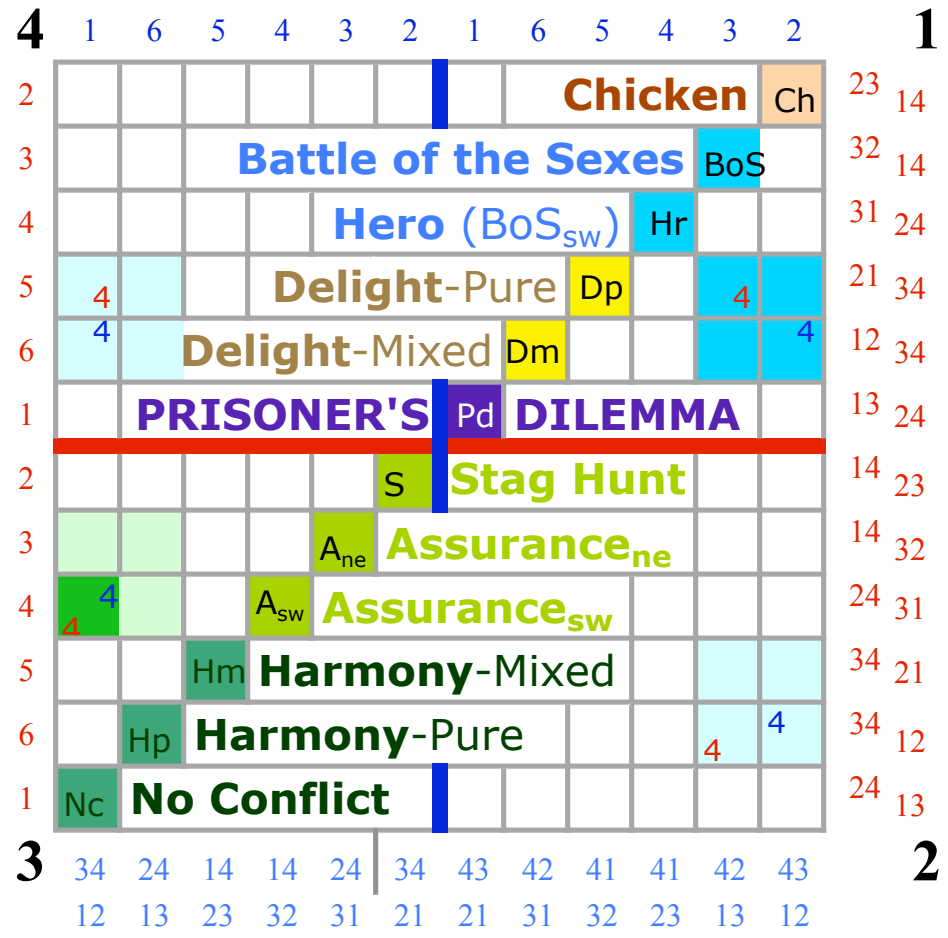

Row payoffs same across row, column same down

## b. Dominant Strategies and Nash Equilibria

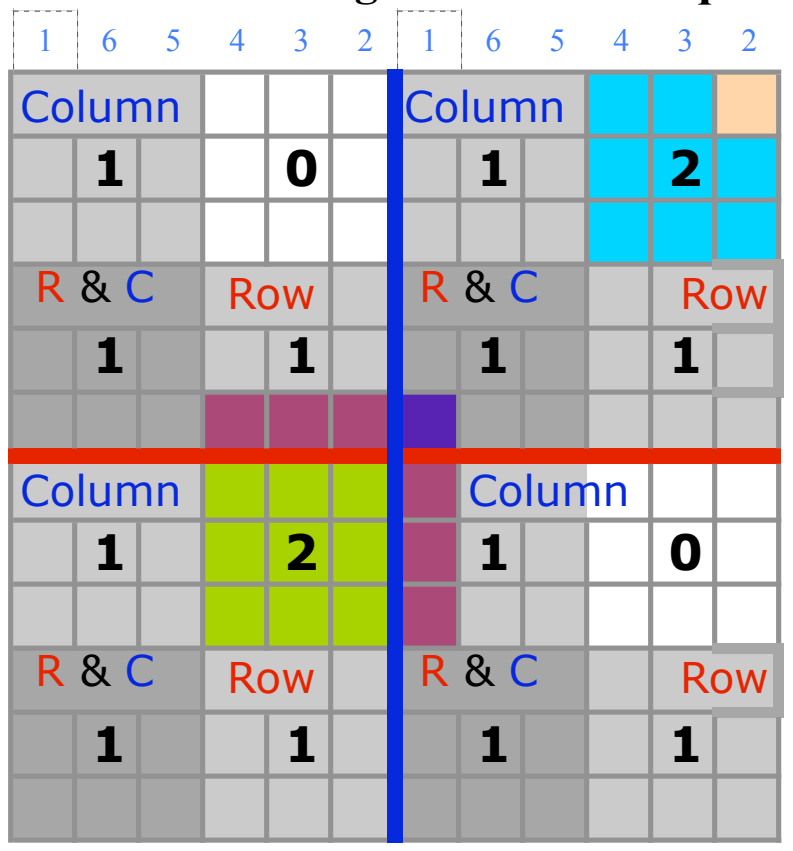

Row has dominant strategies in lower 3 rows, column in left 3 columns, of each layer

Areas with 0, 1, or 2 Nash Equilbria;

## c. Payoff Familes and Subfamilies

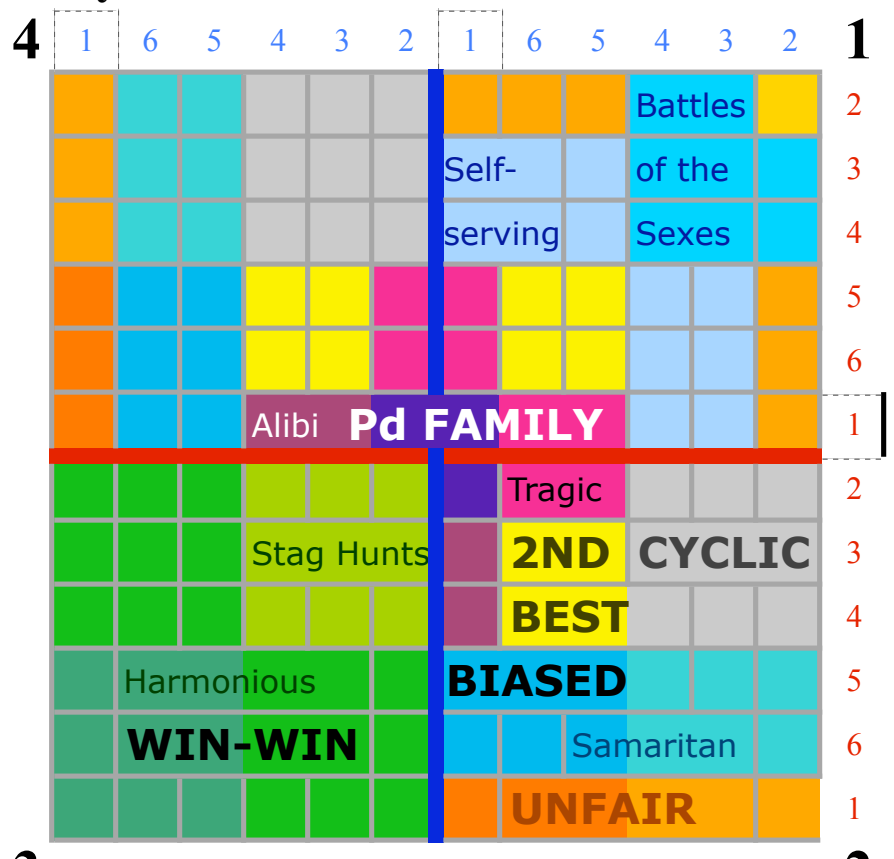

3 Tragic subfamily added to Pd family 2

Added families: Second Best, Biased, and Unfair

Subfamilies: Samaritan, Self-serving, & Harmonious

## d. High swaps (3↔4) link layers

6 hotspots double-link tiles on two layers by 3↔4 swaps

6 pipes link 4 tiles on 4 layers

3↔4 swap layer vectors — X |

Inside tile: ↕ row swaps row. ↔ column swaps column

4 Scrolled to natural layout: Pd neNEne 1

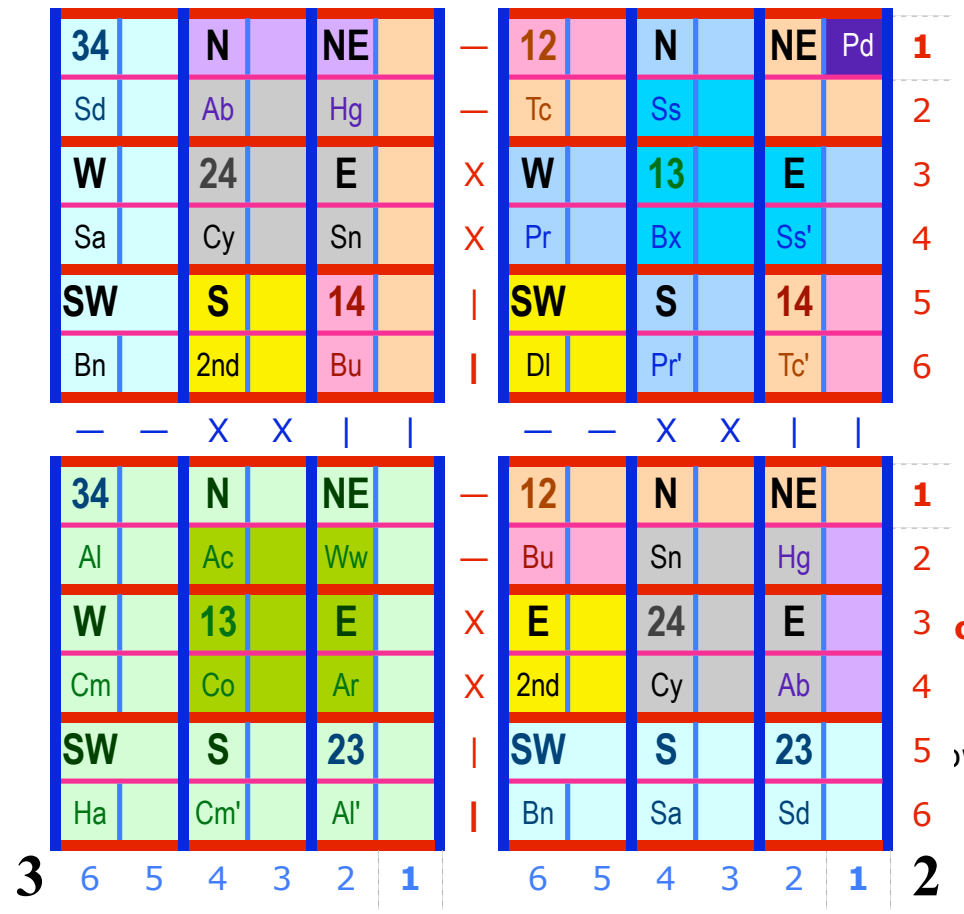

## e. Order Diagrams Show Symmetries

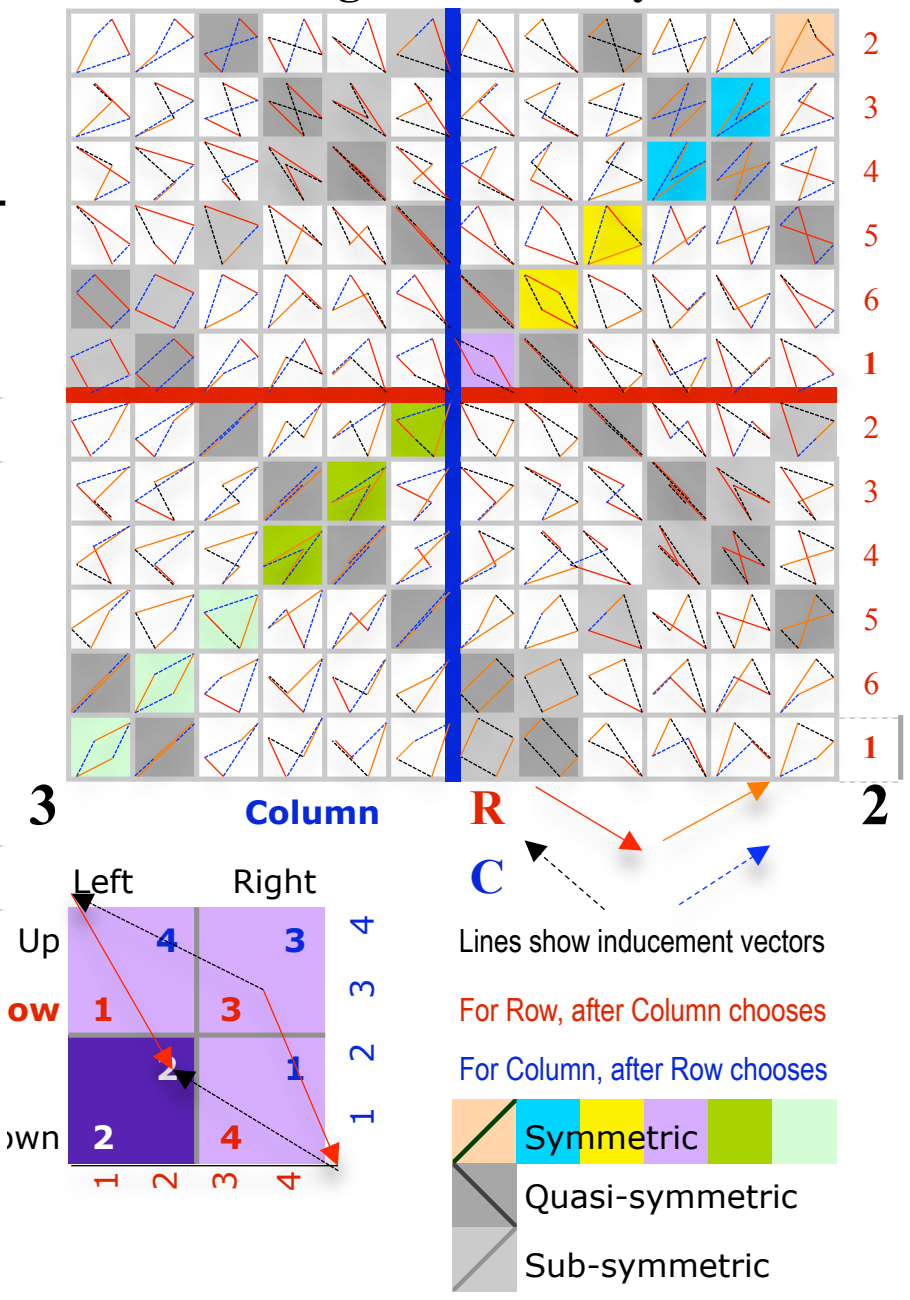

## f. Interests Aligned or Mixed

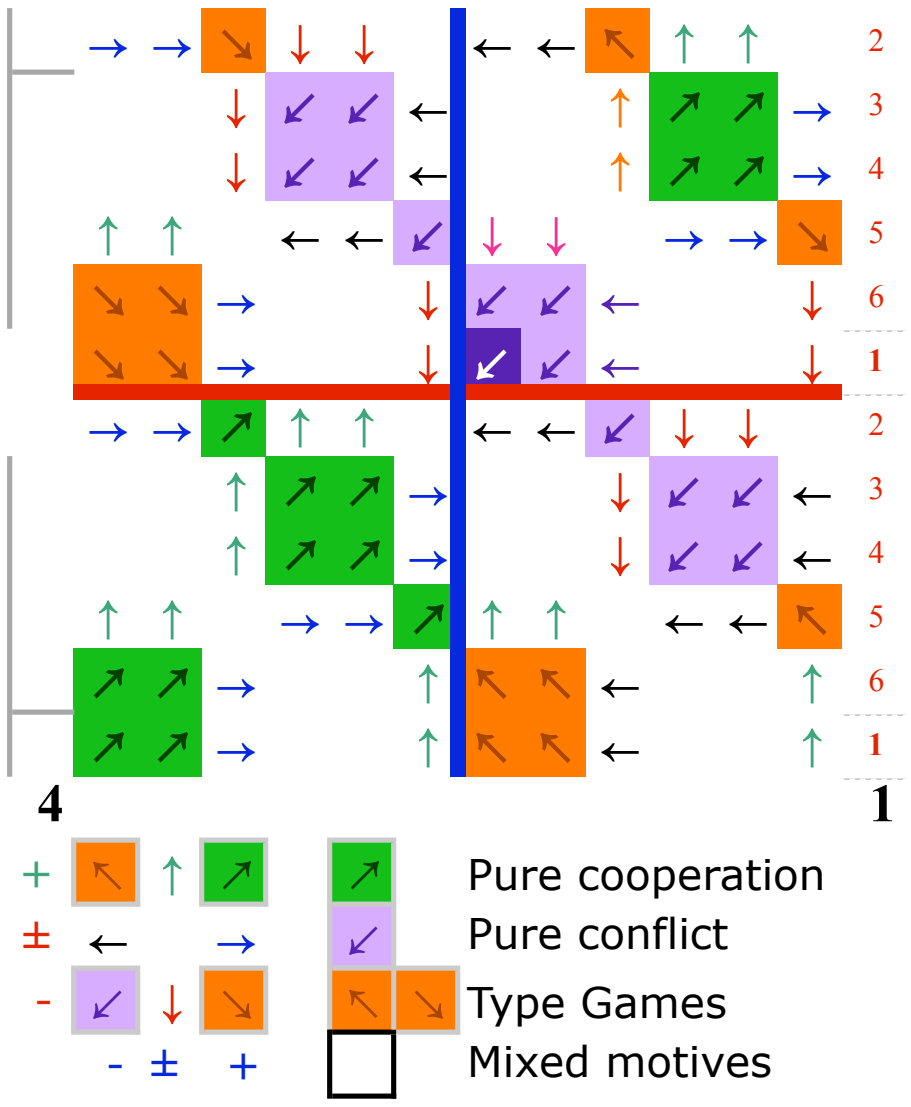

For inducement correspondences, see: Greenberg 1990 *The Theory of Social Situations*

For Type games, see Robinson & Goforth 2005

## g. Games with Ties are within the Topology

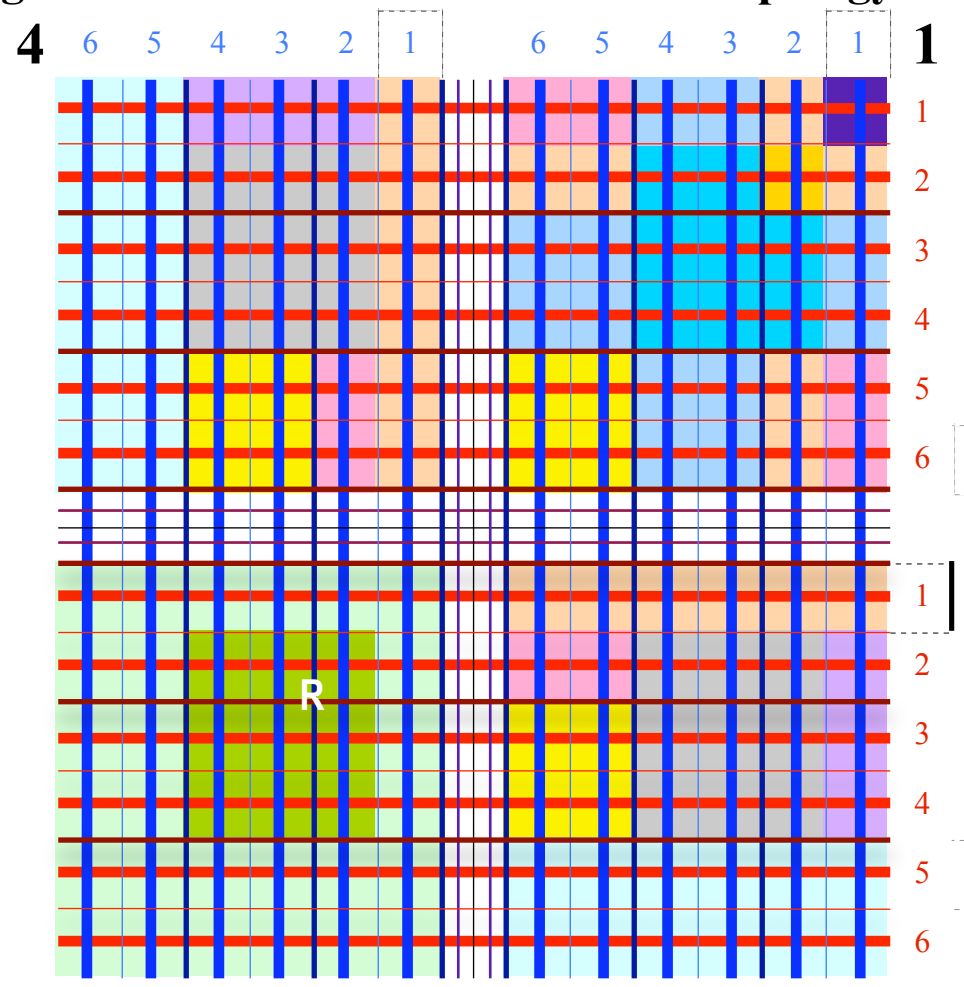

3 Swaps1↔2&2↔3 Half-swaps 1=2&2=3 PdneNl 2

Ordinal games are at intersections (nodes/vertices)

Games with ties (non-strict) lie between strict ordinal games (as do other normalized games)

See Plate 3b and see Robinson, Goforth & Cargill 2007

## h. Rapoport, Guyer & Gordon Taxonomy

| | 1 | 6 | 5 | 4 | 3 | 2 | 1 | 6 | 5 | 4 | 3 | 2 |
|---|---|---|---|---|---|---|---|---|---|---|---|---|
| 2 | 55 | 50 | 49 | 70 | 78 | 72 | 39 | 35 | 36 | 65 | 67 | 66 |
| 3 | 56 | 52 | 51 | 74 | 76 | 71 | 37 | 31 | 32 | 64 | 68 | 67 |
| 4 | 44 | 41 | 40 | 73 | 75 | 77 | 38 | 33 | 34 | 69 | 64 | 65 |
| 5 | 18 | 16 | 15 | 53 | 42 | 45 | 10 | 8 | 7 | 34 | 32 | 36 |
| 6 | 17 | 14 | 13 | 54 | 43 | 46 | 11 | 9 | 8 | 33 | 31 | 35 |
| 1 | 21 | 19 | 20 | 57 | 47 | 48 | 12 | 11 | 10 | 38 | 37 | 39 |
| 2 | 26 | 22 | 23 | 58 | 62 | 61 | 48 | 46 | 45 | 77 | 71 | 72 |
| 3 | 27 | 24 | 25 | 59 | 63 | 62 | 47 | 43 | 42 | 75 | 76 | 78 |
| 4 | 30 | 28 | 29 | 60 | 59 | 58 | 57 | 54 | 53 | 73 | 74 | 70 |
| 5 | 2 | 4 | 5 | 29 | 25 | 23 | 20 | 13 | 15 | 40 | 51 | 49 |
| 6 | 1 | 3 | 4 | 28 | 24 | 22 | 19 | 14 | 16 | 41 | 52 | 50 |
| 1 | 6 | 1 | 2 | 30 | 27 | 26 | 2 | 17 | 18 | 44 | 56 | 55 |

| | |
|---|---|
| No conflict | Strongly stable |
| No equilibria | Stable |
| 2 equilibria, non-eq. solution | Force-vulnerable |
| 2 equilibria, eq. solution | Threat-vulnerable |
| Strongly stable deficient | Unstable |

Adapted from Robinson & Goforth 2003; *see* Rapoport et al. 1976 *The 2x2 Games*

## i. Brams Typology and Game Numbers

| | 1 | 6 | 5 | 4 | 3 | 2 | 1 | 6 | 5 | 4 | 3 | 2 |
|---|---|---|---|---|---|---|---|---|---|---|---|---|
| 2 | 50 | 37 | 36 | 46 | 31 | 29 | 22 | 18 | 19 | 52 | 53 | 57 |
| 3 | 56 | 39 | 38 | 43 | 45 | 47 | 20 | 14 | 15 | 51 | 54 | 53 |
| 4 | 49 | 13 | 12 | 42 | 44 | 30 | 21 | 16 | 17 | 55 | 51 | 52 |
| 5 | 6 | 4 | 3 | 40 | 23 | 25 | 10 | 8 | 7 | 17 | 15 | 19 |
| 6 | 5 | 2 | 1 | 41 | 24 | 26 | 11 | 9 | 8 | 16 | 14 | 18 |
| 1 | 35 | 33 | 34 | 48 | 27 | 28 | 32 | 11 | 10 | 21 | 20 | 22 |
| 2 | No numbers for "no conflict" games | | | | | | 28 | 26 | 25 | 30 | 47 | 29 |
| 3 | | | | | | | 27 | 24 | 23 | 44 | 45 | 31 |
| 4 | | | | | | | 48 | 41 | 40 | 42 | 43 | 46 |
| 5 | | | | | | | 34 | 1 | 3 | 12 | 38 | 36 |
| 6 | | | | | | | 33 | 2 | 4 | 13 | 39 | 37 |
| 1 | | | | | | | 35 | 5 | 6 | 49 | 56 | 50 |

| | |
|---|---|
| Non-myopic equilibria | Not cyclic |
| NMEs 3 2 1 | Weakly cyclic |
| See Brams 1994 | Moderately cyclic |
| *Theory of Moves* | Strongly cyclic |

# 3. Ties Make Simpler Games Within the Topology of 2x2 Games

## a. Ties Make Games in Between

Chicken's Dilemma at the center of the Pd-Chicken Tile

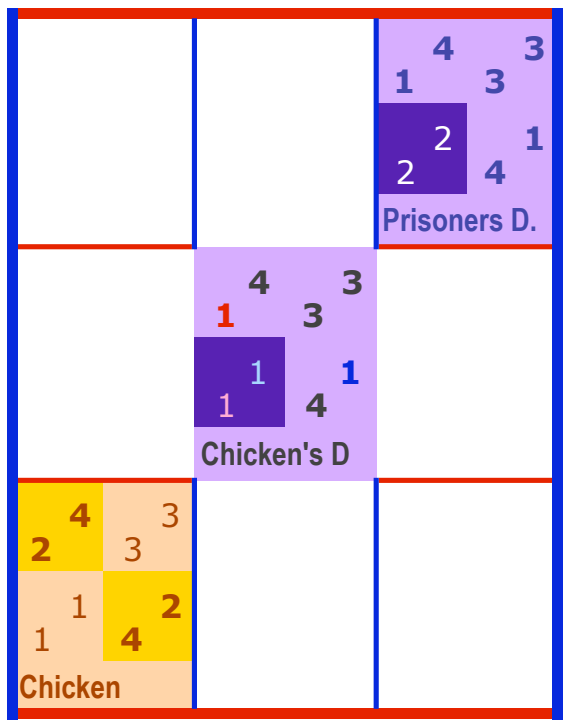

## b. A Herd of Stag Hunts

Rouseau's Stag and Hare Hunt diagonally between tiles. 2=3 half-swaps

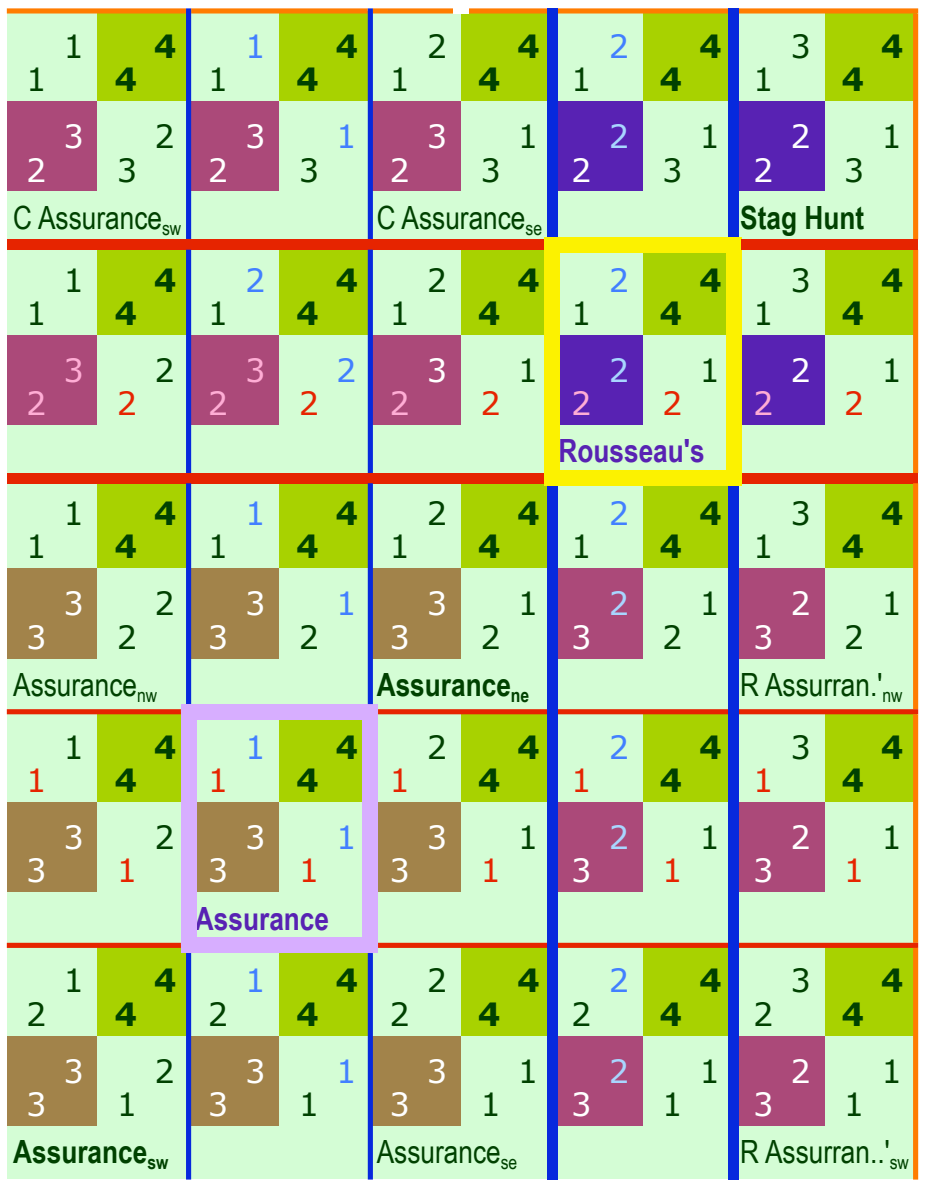

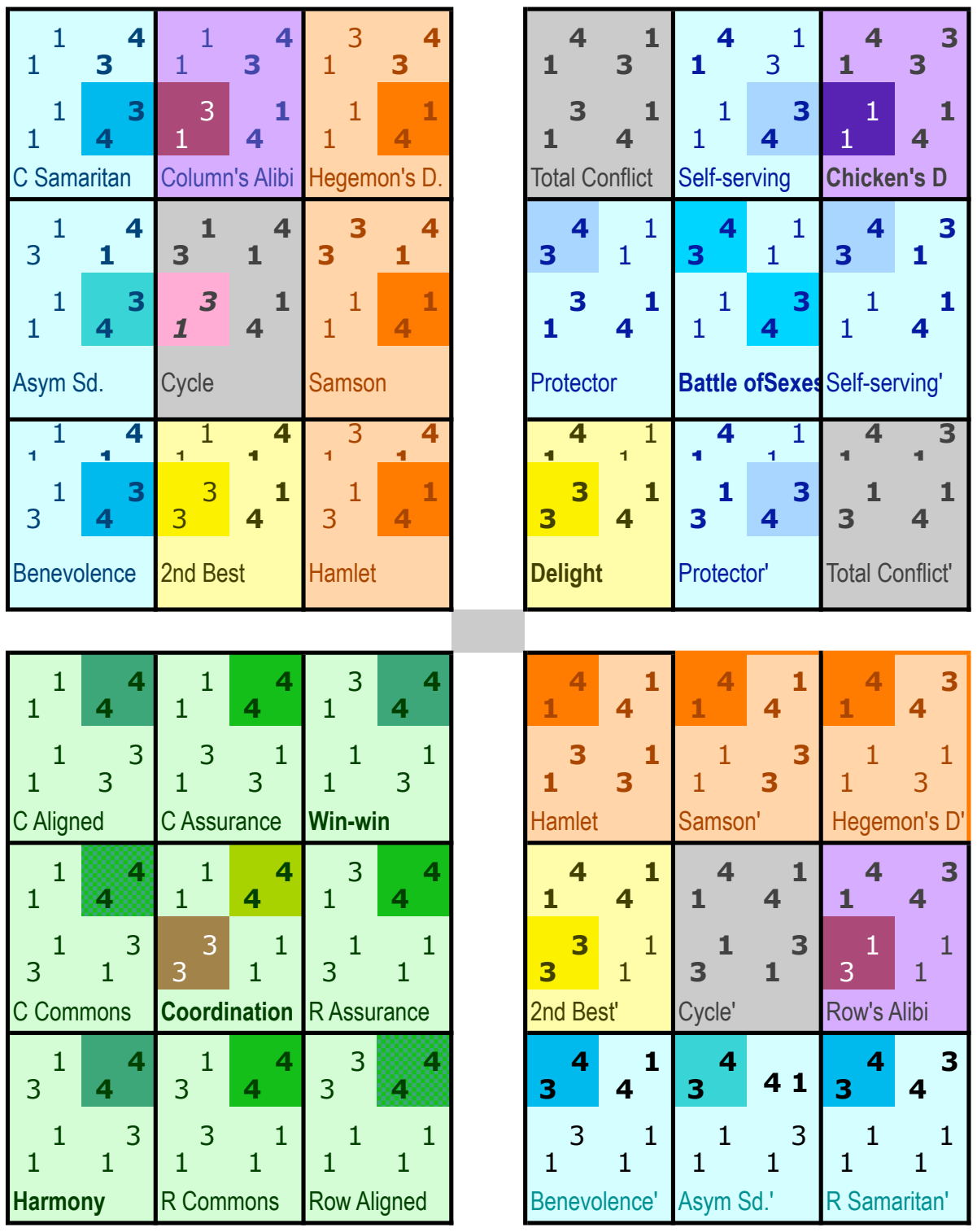

## c. Tile Games

Made by two 1=2 half-swaps

$2_3$ 0111

E 1, 2=3=4

## d. Simple Games

Two preferences for both players, one, two, or three ties

$2_2$ 0011

C 1=2, 3=4

B, C, E Distinct Payoff Classes

see Fraser & Kilgour 1988

1=2 Half-swap payoff tie or split

see Robinson, Goforth & Cargill 2007:9

Unshaded games are equivalents by swaps of rows, columns or both

$2_1$ 0001

B 1=2=3, 4

Win-win equilibrium

Win-Lose

No dominance

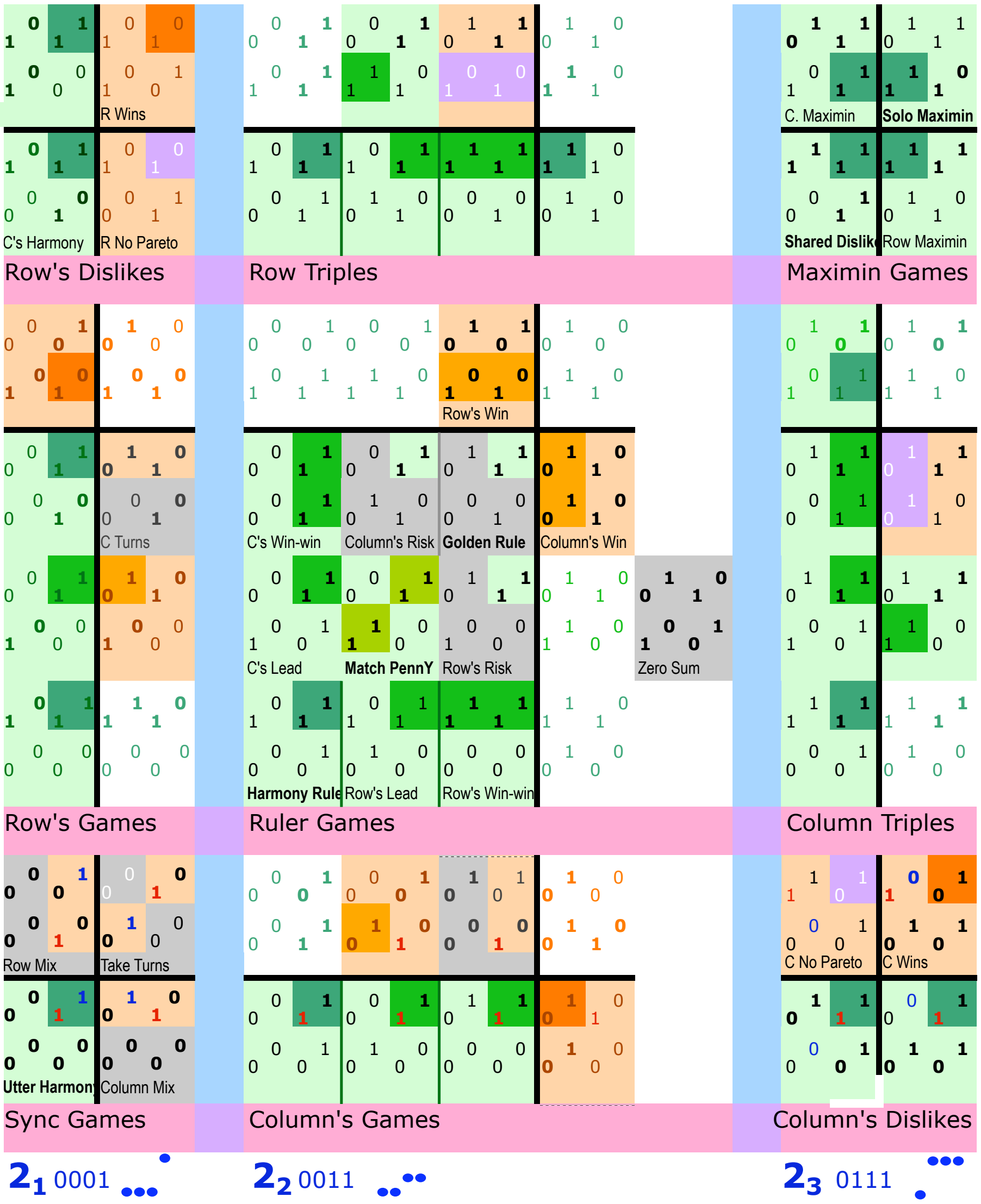

# 4. Navigation Chart for the Topology of 2x2 Ordinal Games with Ties

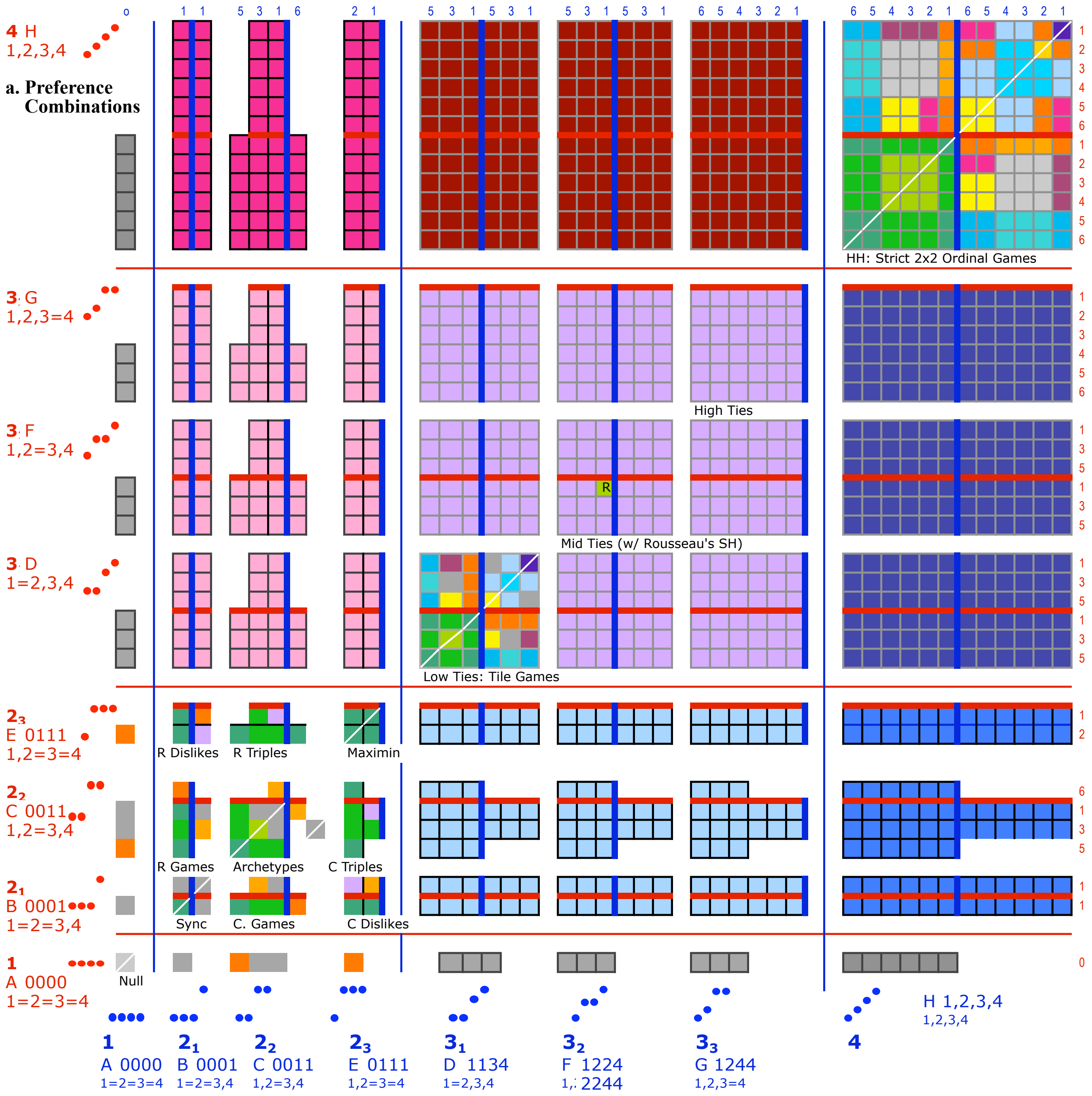

## b. Ordinal Games Categorized by Distinct Payoff Classes

| | | | | | | | | | |
|---|---|---|---|---|---|---|---|---|---|
| 4 | H | 6 | 24 | 36 | 24 | 72 | 72 | 72 | **144** |
| $3_3$ | G | 3 | 12 | 18 | 12 | 36 | 36 | 36 | 72 |
| $3_2$ | F | 3 | 12 | 18 | 12 | 36 | 36 | 36 | 72 |
| $3_1$ | D | 3 | 12 | 18 | 12 | **36** | 36 | 36 | 72 |
| $2_3$ | E | **1** | **4** | **6** | **4** | 12 | 12 | 12 | 24 |
| $2_2$ | C | **3** | **6** | **<u>12</u>** | **6** | 18 | 18 | 18 | 36 |
| $2_1$ | B | **1** | **4** | **6** | **4** | 12 | 12 | 12 | 24 |
| 1 | A | **1** | **1** | **3** | **1** | 3 | 3 | 3 | 6 |
| Total 1413 | | A | B | C | E | D | F | G | H |
| | | **1** | **$2_1$** | **$2_2$** | **$2_3$** | **$3_1$** | **$3_2$** | **$3_3$** | **4** |

## c. Half-swap Pathways between All ties and no ties

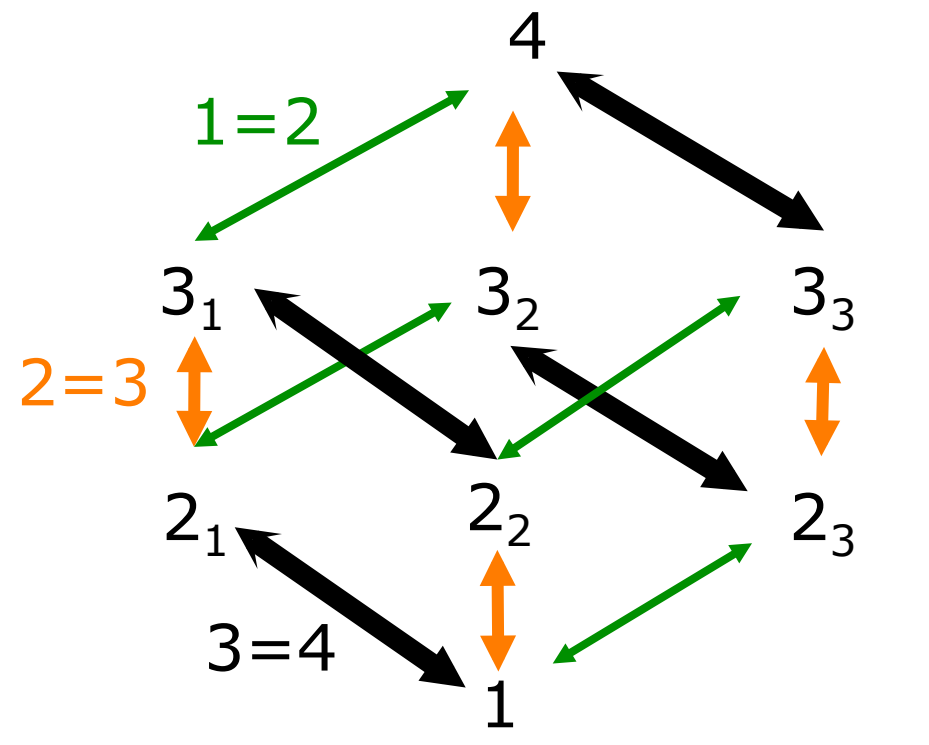

## d. Beyond Extending the Topology

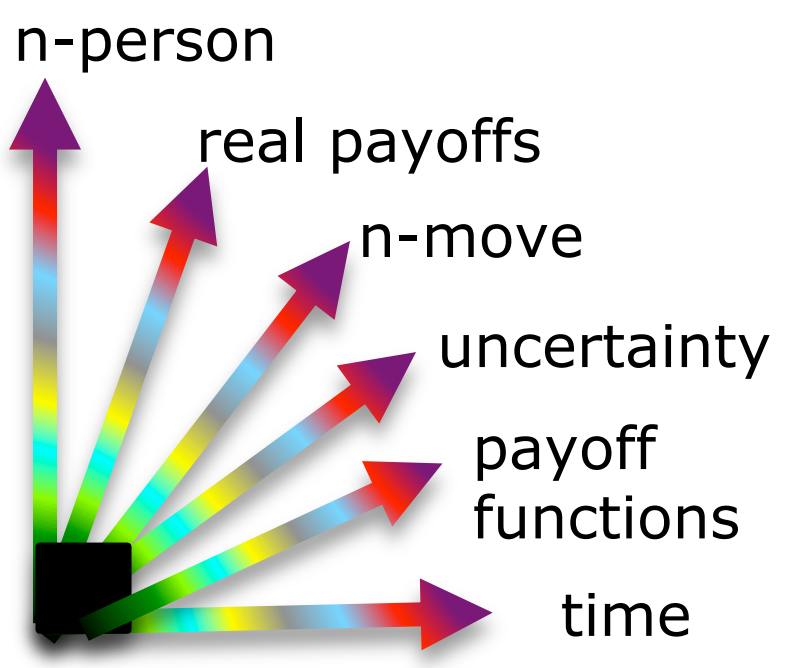

Adapted from Robinson, Goforth & Cargill 2007:9, 13; see Fraser and Kilgour 1988